\documentclass[runningheads,a4paper]{llncs}

\usepackage{graphicx}        
\usepackage{amsmath}
\usepackage{amsfonts}
\usepackage{amssymb}
\usepackage{algorithm2e}
\usepackage{url}
\usepackage{cite}

\urldef{\mailsa}\path|{Vibeke.Skytt, Tor.Dokken, Heidi.Dahl}@sintef.no|
\urldef{\mailsb}\path|{Q.Harpham}@hrwallingford.com|

\begin{document}
\mainmatter
\title{Deconfliction and Surface Generation from Bathymetry Data Using LR B-splines}
\titlerunning{Bathymetry, deconfiction, surface generation and LR B-splines}
\author{Vibeke Skytt$^1$ \and Quillon Harpham$^2$ \and Tor Dokken$^1$ \and
Heidi E.I. Dahl$^1$}
\authorrunning{V. Skytt, Q. Harpham, T. Dokken, H.E.I. Dahl}
\institute{SINTEF, Forskningsveien 1, 0314 Oslo, Norway$^1$ \\
HR Wallingford, Howbery Park, Wallingford, Oxfordshire 0x10 8BA, 
United Kingdom$^2$}
\maketitle
\mailsa \\
\mailsb \\

\begin{abstract}
A set of bathymetry point clouds 
acquired by different measurement techniques at 
different times, having different accuracy and varying patterns of points,
are approximated by an LR B-spline surface. The aim is 
to represent the sea bottom with good accuracy 
and at the same time reduce the data size considerably. In this process the
point clouds must be cleaned by selecting the ``best'' points for surface
generation. This cleaning process is called deconfliction, and we
use a rough approximation of the combined point clouds as a reference surface 
to select a consistent set of points.
The reference surface is updated with the selected points to create an accurate 
approximation. LR B-splines is the selected surface format due to its suitability for adaptive refinement and approximation, and its ability 
to represent local detail without a global increase in the data size of the 
surface. 
\end{abstract}

\section{Introduction}
Bathymetry data is usually obtained by single or multi beam sonar or
bathymetry LIDAR.
Sonar systems acquire data points by collecting information from reflected 
acoustic signals. 
Single beam sonar is the traditional technique for
acquiring bathymetry data and it 
collects data as discrete point data long the the path of
a vessel equipped with single beam acoustic depth sounders. The equipment is
easy to attach to the boat and the acquisition cost is lower than for
alternative acquisition methods. The obtained data sets, however, have a scan
line like pattern, which gives a highly inhomogeneous point cloud as input to a
surface generation application. 

Acquisition of bathymetric data with Multi Beam Echo Sounder (MBES) is 
nowadays of common use. A swath MBES system produces multiple acoustic beams 
from a single transducer in a wide angle. It generates
points in a large band around the vessel on which the equipment is installed.
The swath width varies from 3 to 7 times the water depth. 
In shallow areas, the results of a multi beam sonar degenerates to that of
the single beam sonar as the sonar angle is reduced due to a short distance
to the sea bottom. Multi beam sonar data acquisition is described in 
some detail in~\cite{outliers4}.

LIDAR (light detection and ranging) measures elevation or depth by analyzing
the reflections of pulses of laser light from an object.
Near shore, especially in shallow areas or in rough waters that are
difficult to reach by a sea-born vessel, data acquisition using 
bathymetry LIDAR is a good alternative to sonar.
Bathymetry LIDAR differs from topography LIDAR by the wavelength of the signals
that are used. To be able to penetrate the water, a shorter wavelength is
required, so green light is used instead of red. This change
reduces the effect of the power used by the laser, and bathymetry 
LIDAR becomes more costly than the topography equivalent. 

Our aim is to represent a specified region with a seamless surface. 
Some parts of the region are only covered by one survey, while other areas are 
covered by numerous surveys obtained by different
acquisition methods. Where no survey data exists,
even vector data created from navigation charts may be taken as input.
Collections of bathymetric surveys are a source of potentially ``big data'' 
structured as point clouds. Individual surveys vary both spatially 
and temporally and can overlap with many other similar surveys. Where depth 
soundings differ greatly between surveys, a strategy needs to
be employed to determine how to create an optimal bathymetric surface based on 
all of the relevant, available data, i.e., select the best data for surface
creation.

The digital elevation model (DEM) is the most common format for representing
surfaces in geographical information systems (GIS). DEM uses a raster format 
for storage. Rasters are rectangular arrays of cells (or pixels), each of 
which stores a value for the part of the surface it covers. A given cell 
contains a single value, so the amount
of detail that can be represented for the surface is limited by the
raster cell resolution. 
The elevation in a cell is frequently estimated using the height
values of nearby points. The estimation
methods include, but are not restricted to, the inverse weighted
interpolation method, also called Shepard's method~\cite{grid:shepard},
natural neighbour interpolation, radial basis functions and kriging 
~\cite{rbf1, grid:interpolate, grid:kriging}.
Alternatively, one of the existing points lying within the cell can be
selected to represent the cell elevation.

Triangulated irregular network (TIN) is
used to some extend in GIS context. Sample data points serve as
vertices in the triangulation, which normally is computed as a 
Delaunay triangulation.
A triangulated surface can interpolate all points
in the point cloud exactly, 
but for large data sizes an approximate solution is more
appropriate. The triangulation data structure is flexible and an 
irregular and well-chosen
distribution of nodes allows capturing rapid changes in the represented
sea bed or terrain.

The purpose of trend surfaces is not representation of terrains, 
but data analytics. These surfaces are described by polynomials of low degree
globally approximating the data. 
Trend surface analysis is used to identify general trends in the data and the 
input data can be separated into two components: the trend corresponding 
to the concept of regional features and the residual corresponding to 
local features. Very often, however, the global, polynomial surface becomes 
too simplistic compared to the data.

In GIS context, splines are almost entirely understood as regularized splines
or splines in tension in the context of radial basis functions. 
Only in rare instances splines are used for terrain modeling. 
However, Sulebak et. al.,~\cite{Sulebak}, use multi-resolution splines 
in geomorphology. We aim at using polynomial spline surfaces to represent our
final result. Moreover, in the process of selecting data surveys for the
surface generation, we use spline surfaces as extended trend surfaces.
Spline surfaces are able to compactly represent smooth shapes, but our
bathymetry data are not likely to describe a globally smooth seabed. Thus,
we turn our attention towards locally refineable splines in the form of
LR B-spline surfaces.

Section~\ref{LRsplines} gives a brief overview of the concept of LR B-splines.
In Section~\ref{surfgen}, we will present the construction of LR B-spline
surfaces and collections of such surfaces approximating point clouds from 
bathymetry data. The topic of Section~\ref{deconfliction} is the deconfliction
process discussed in the context of outliers detection, both
for Geo-spatial data and in a more general setting. Finally, we will present
a conclusion including plans for further work in Section~\ref{conclusion}.

\section{LR B-splines} \label{LRsplines}
\begin{figure}
\begin{center}
\includegraphics[width=6.5cm]{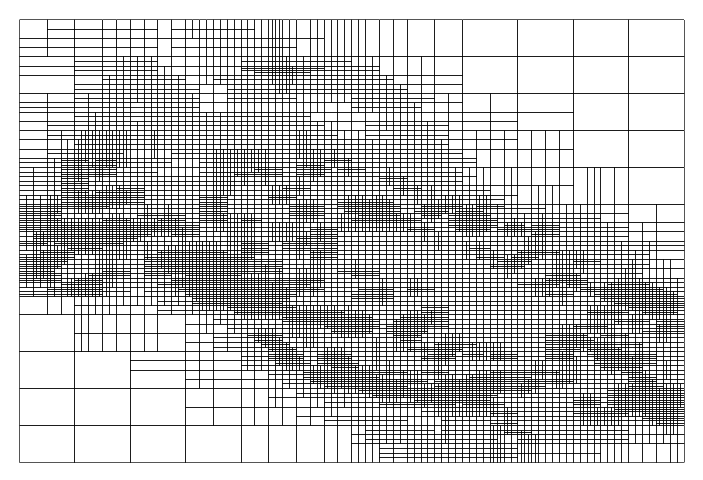}
\end{center}
\caption{The polynomial patches in the domain of an LR B-spline surface. This
construction will be discussed in some detail in Section~\ref{surfgenexample}.
\label{fig:box_partition} }
\end{figure}
LR B-spline surfaces are spline surfaces defined on a box partition as
visualized in Figure~\ref{fig:box_partition}, see~\cite{lr:lrsplines} for
a detailed description of the theory.

In contrast to the well-known tensor-product spline surfaces, LR B-spline spline
surfaces posses the property of local refineability. New knot lines, 
{\it not} covering the entire domain of the surface, can be added to the surface
description. The new knot line must, however, cover the support of at least one
B-spline. The local refinement property implies that models with varying
degree of detail can be represented without the drastic increase in model size
that would arise in the tensor-product representation.
Other approaches 
addressing the problem of lack of local refimenent methods in the tensor-product
construction are hierarchical splines~\cite{approx:hierarchical} 
and T-splines~\cite{lr:tsplines}.

An LR-B spline surface $F$ is expressed with respect to parameters $u$ and $v$ as 
\[
F(u,v) = \sum_{i=1}^L s_i P_i N_i^{d_1,d_2}(u,v),
\]
where $P_i$ are the surface coefficients, $N_i$ are the associated B-splines and $s_i$ are scaling factors that ensure partition of unity. 
The B-splines are constructed by taking the tensor-products of univariate B-splines, and are thus defined on a 
set of knots in both parameter directions. They have polynomial degree $d_1$ and $d_2$ in the first and second parameter direction, respectively. 

LR B-spline surfaces possess most of the properties of tensor-product spline surfaces, such as non-negative 
B-spline functions, limited support of B-splines and partition of unity, 
which ensure numerical stability and modelling accuracy. 
Linear independence of the 
B-spline functions is not guaranteed by default. For
LR B-spline surfaces of degree two and three 
and knot insertion restricted to the middle of knot intervals, no cases of
linear dependency are known, but the mathematocal proof is still 
not completed. Actual
occurrences of linear dependence can be detected by the peeling algorithm,
~\cite{peeling}, and it can be resolved by a strategy of carefully 
chosen knot insertions.

\section{Surface Generation} \label{surfgen}

We assume the input to be one point cloud where the
initial bathymetry data is translated to points represented by their $x$, $y$, 
and $z$-coordinates. The points can be obtained from one data survey or
collected from several surveys. No further preprocessing of the 
points is performed. 

To exploit the local refineability of the LR B-spline surfaces and to
optimize the positioning of the degrees of freedom in the surface, we
apply an adaptive surface generation approach using two different 
approximation methods over given spline spaces.

Due to the acquisition methods, bathymetry data is normally projective onto
their $x$ and $y$-coordinates. Thus,
it is possible to parameterize the points by these coordinates
and approximate the height values ($z$-coordinates) by a function. In
steep areas, however, a parametric surface would be more appropriate. This issue is
discussed in~\cite{IQmulusbook}. In this paper, we will concentrate on 
approximation of height values.

The description of the surface generation method in the remainder of this 
section is partly fetched from~\cite{LRapprox} and~\cite{IQmulusbook}.
\subsection{An Iterative Framework for Approximation with LR-spline Surfaces} \label{sec:adaptive}

The aim of the approximation is to fit an LR-spline surface to a given point cloud within a certain threshold or tolerance.  
Normally this is achieved for the majority of points in the cloud, and any remaining points that are not within the tolerance after a certain number of iterations can be subject to further investigation.
Algorithm~\ref{alg:framework} outlines the framework of the adaptive
surface approximation method.

\begin{algorithm}
\KwData{input point cloud, parameters governing the adaptive procedure: 
tolerance and maximum number of iterations}
\KwResult{LR B-spline surface and accuracy information(optionally)}
Initiate LR/tensor-product space\;
Generate initial surface approximation\;
\While{there exist out-of-tolerance points or max-levels not reached} {
 \For{points within each polynomial patch} {
  Compute the max. error between points and surface\;
  \If{max. error is greater than tolerance} {
   Refine LR B-spline surface\;
  }
 }
 Perform an iteration of the chosen approximation algorithm\;
}
\caption{The LR B-spline surface generation algorithm}
\label{alg:framework}
\end{algorithm}

The polynomial bi-degree of the generated LR B-spline surface can be of any degree higher
than one, however, in most cases a quadratic (degree two) surface will suffice.
Quadratic surfaces ensure $C^1$-continuity across knot lines with multiplicity one, and 
as terrains often exhibits rapid variations higher order smoothness may be too restrictive.

The algorithm is initiated by creating a coarse tensor-product spline space.
An initial LR B-spline surface is constructed by
approximating the point cloud in this spline space. A tensor-product spline 
space can always be represented by an LR B-spline surface while an LR B-spline
surface can be turned into a tensor-product spline surface by extending
all knot lines to become global in the parameter domain of the surface.

In each iteration step, a surface approximation is performed.
Two approximation methods are used for this purpose, least squares
approximation  and multi-resolution B-spline approximation (MBA). 
Both approximation methods are general algorithms 
applied to parametric surfaces, which have been adapted for use with LR B-splines.
Typically least squares approximation is used for the first iterations
as it is a global method with very good approximation properties, while we
turn to the MBA method when there is a large variety in the size of 
the polynomial elements of the surface. A comparison of the performance 
of the two methods can be found in~\cite{LRapprox}. 
The distances between the points in the point cloud 
and the surface is computed to produce a distance field. In our setting the
surface is parameterized by the $xy$-plane 
and the computation can be performed by a 
vertical projection mainly consisting of a surface evaluation.

Next we identify the regions of the domain that do not
meet the tolerance requirements and refine the representation in these areas
to provide more degrees of freedom for the approximation. 
Specifically, we identify B-splines whose support contain data points where the accuracy 
is not satisfied, in their support are identified and introduce 
 new knot lines, in one or two
parameter directions depending on the current distance field configuration.
The new knot lines must cover the support of at least one
B-spline.
In each iteration step, many new knot line segments will be inserted
in the surface description, giving rise to the splitting of many B-splines.
The splitting of one B-spline may imply that an existing 
knot line segment partly covering its support will now completely
cover the support of one of the new B-splines that, in turn, is split
by this knot line. 

\subsection{Least Squares Approximation}\label{sec:lrls}
Least squares approximation is a global method for surface approximation where 
the following penalty function is minimized with respect to the 
coefficients $P_i,$ over the surface domain, $\Omega$:
\[
\alpha_1 J(F) + \alpha_2\sum_{k=1}^K (F(x_k,y_k)-z_k)^2.
\]
Here $\mathbf{x}_k = (x_k,y_k,z_k), k=1,\ldots,K$, are the input data points. 
$J(F)$ is a smoothing term, which is added to the functional
to improve the surface quality and ensure a solvable system
even if some basis functions lack data points in their support.
The approximation is weighted (by the scalars $\alpha_1$ and $\alpha_2$) in order to favour either the smoothing term or the least squares approximation, respectively. 
The smoothing term is given by 
\begin{equation}
J(F) = 
\iint_\Omega \int_0^\pi \sum_{i=1}^3 w_i \left(\frac{\partial^iF(x+r\cos\phi,y+r\sin\phi)}{\partial r^i}\bigg|_{r=0}\right) \text{ d}\phi\text{d}x\text{d}y.
\label{eq:Jf}
\end{equation}

The expression approximates the minimization of a measure involving surface 
area, curvature and variation
in curvature. Using parameter dependent measures, the minimization
of the approximation functional is reduced to solving a linear equation system.
In most cases $w_1=0$ while $w_2=w_3$. In our case, however, $w_2=1$ and 
$w_3=0$ as we utilize 2nd degree polynomials.
A number of smoothing terms exist. The one given above is presented 
in~\cite{spline:smooth1}. Other measures can be found
in~\cite{spline:smooth2}, and~\cite{approx:greiner} looks into the effect 
of choosing different smoothing functionals.

In Equation~\ref{eq:Jf}, a directional derivative is defined from the first,
second and third derivatives of the surface, and
in each point $(x,y)$ in the parameter domain,
this derivative is integrated radially. The result is integrated over 
the parameter domain. 

Experience shows that the approximation term must be 
prioritized in order to achieve a good approximation to the data points. 
This is in conflict with the role of the
smoothing term as a guarantee for a solvable equation system. 
Estimated height values
in areas sparsely populated by data points, are thus included to stabilize the
computations. Some details on the stability of least squares
approximation used in this context can be found in~\cite{LRapprox}.

\subsection{Locally Refined Multilevel B-spline Approximation (LR-MBA)}\label{sec:lrmba}

Multilevel B-spline approximation (MBA) is a local approximation method~\cite{lr:mba}.
Surface coefficients are computed with respect to the distances between 
data points in the support of the B-spline functions corresponding to the 
coefficients and a current surface. 
The procedure is explicit and does not require solving an equation system.

A surface approximating the residuals between a point cloud and a current
surface is computed as follows.
Let $\mathbf{x}_c = (x_c,y_c,z_c), c=1,\ldots,C$, be the data points above 
the support of a given B-spline and $r_c$ the residual corresponding
to $\mathbf{x}_c$. As the
initial point cloud is scattered, there is a large variation in the number
of points. If the
B-spline has no points in its support or if all the points are closer to the surface than a prescribed
tolerance, the corresponding coefficient is set to zero. Otherwise,
a coefficient $P_i$ is determined by the following expression:
\[
P_i = \frac{\sum_{c} (s_i N_i(x_c,y_c))^2\phi_c}{\sum_{c} (s_iN_i(x_c,y_c))^2},
\]
where $\phi_c$ is computed for the residual of each data point as 
\[
\phi_c = \frac{s_iN_i(x_c,y_c)r_c}{\sum_l (s_lN_l(x_c,y_c))^2}.
\]
The sum in the denominator is taken over all B-splines which contain $(x_c,y_c)$ in their support.

The algorithm is based on a B-spline approximation technique proposed for image morphing and is 
explained in~\cite{lr:mba2} for multilevel B-spline approximation. 
In the original setting a number of difference surfaces approximating the 
distances between the point cloud and the current surface is computed. 
The final surface is evaluated by computing the sum of the initial 
surface and all the difference surfaces. In the LR B-splines setting, the 
computed difference function is incrementally added to the 
initial surface at each step giving a unified expression for the surface.

\subsection{Tiling and Stitching} \label{tilestitch}
Very large point clouds are unfit for being approximated by one surface due
to memory restrictions and high computation times. During surface generation
each data point is accessed a number of times,
and a tiling approach allows for efficient parallelization over several
nodes. Moreover, a large number of points are potentially 
able to represent a high level of detail, which gives rise to 
approximating LR B-spline surfaces with higher data size. 
The surface size should, however, be restricted as the non-regularity of the polynomial
patches penalizes data structure traversals
when the surface is large (more than 50 000 polynomial patches). 

We apply tiling to improve computational efficiency and limit the size of
the produced surface, and select a regular tiling approach
to enable easy identification
of tiles based on the $x-$ and $y-$ coordinates of the points.
\begin{figure}
\centering
\begin{tabular}{cc}
(a)\includegraphics[width=5cm]{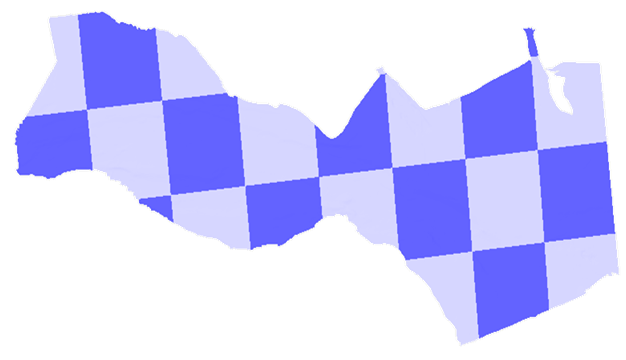}
&(b)\includegraphics[width=5cm]{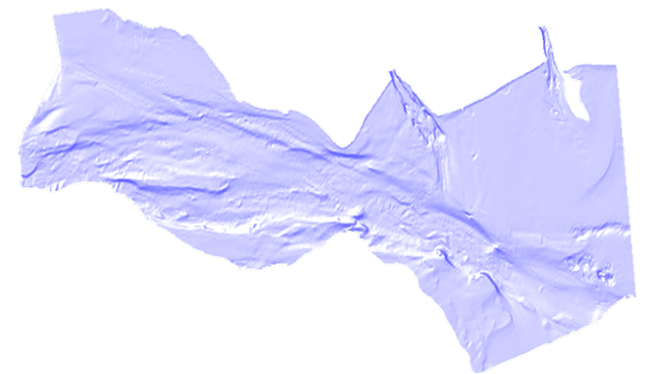}
\end{tabular}
\caption{(a) regular tiling and (b) seamless surface approximating the
tiled data points \label{fig:tile}}     
\end{figure}
Figure~\ref{fig:tile} (a) shows a regular tiling based on a dataset with
131 million points, and (b) a set of LR B-spline surfaces approximating the points.
The computation is done tile by tile, and applying tiles with small overlaps
gives a surface set with overlapping
domains. Each surface is then restricted to the corresponding non-overlapping
tile yielding very small discontinuities between adjacent surfaces.

To achieve exact $C^1$-continuity between the surfaces, stitching is applied.
The surfaces are refined locally along common boundaries to get sufficient
degrees of freedom to enforce the wanted continuity. For $C^0$-continuity 
a common spline space for the boundary curves enables the enforcement of
equality of corresponding coefficients.
$C^1$-continuity is most easily achieved 
by refining the surface to get a tensor-product structure locally along
the boundary and adapting corresponding pairs of coefficients from two adjacent
surfaces along their common boundary to ensure equality of 
cross boundary derivatives.
$C^1$-continuity can always be
achieved in the functional setting, for parametric surfaces it may
be necessary to relax the continuity requirement to $G^1$.

\subsection{Examples} \label{surfgenexample}
\begin{figure}
\begin{center}
\includegraphics[width=5.5cm]{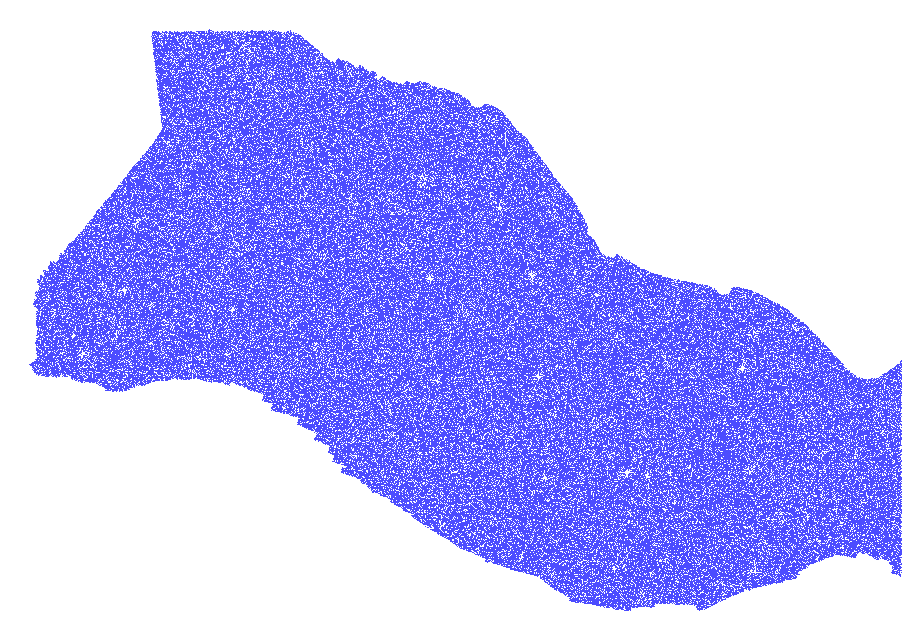}
\end{center}
\caption{Bathymetry point cloud. Data courtesy HR Wallingford: SeaZone \label{fig:pts}}
\end{figure}
{\bf Example 1} We will describe the process of creating an LR B-spline 
surface from a point cloud with 14.6 
million points. The points are stored in a 280 MB binary file. We apply
Algorithm~\ref{alg:framework} using a combination of the
two approximation methods and examine different stages
in the process.
Figure~\ref{fig:pts} shows the point cloud, thinned with a factor of 32 
to be able to distinguish between the points.

\begin{figure}
\centering
\begin{tabular}{ccc}
(a)\includegraphics[width=4.2cm]{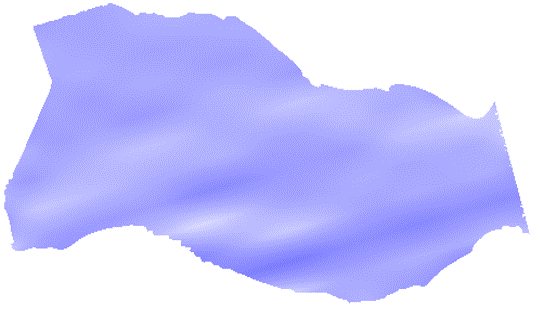}
&(b)\includegraphics[width=3cm]{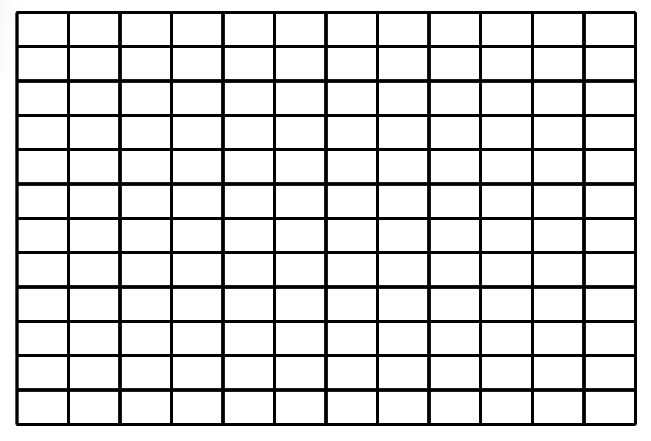}
&(c)\includegraphics[width=3.4cm]{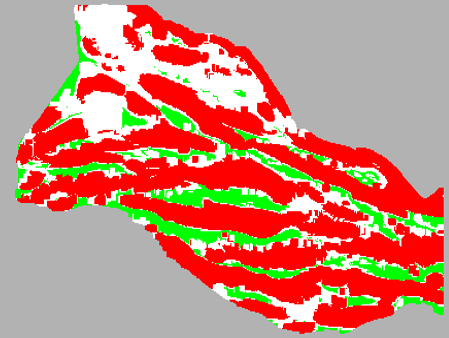}
\end{tabular}
\caption{(a) Initial surface approximation, (b) polynomial patches in the 
parameter domain (element structure) and (c) corresponding distance field. 
White points
lie closer than a threshold of 0.5 meters, red points lie more than 0.5
meters above the surface and green points lie more than 0.5 meters below. \label{fig:sf0}}     
\end{figure}
The initial surface approximation with a lean tensor-product mesh is shown
in Figure~\ref{fig:sf0}. While the point cloud covers a non-rectangular area 
the LR B-spline surface is defined on a regular domain (b), thus the 
surface (a) is
trimmed with respect to the extent of the point cloud. The last figure (c) shows
the points coloured according to the distance to the surface. 
The surface roughly represents
a trend in the point cloud, while the distance field indicates that the
points exhibit a wave-like pattern.

\begin{figure}
\centering
\begin{tabular}{ccc}
(a)\includegraphics[width=4.2cm]{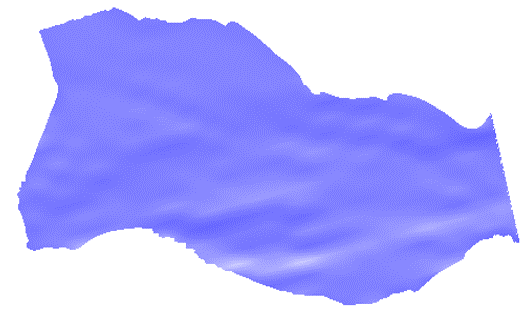}
&(b)\includegraphics[width=3cm]{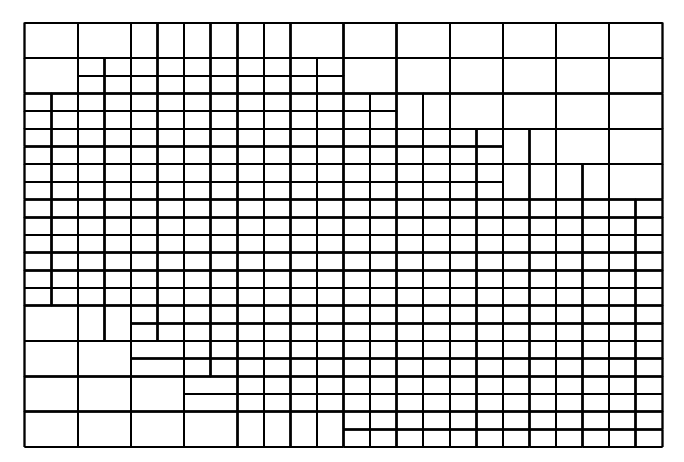}
&(c)\includegraphics[width=3.4cm]{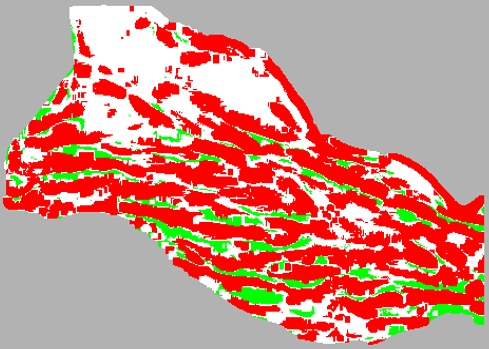}
\end{tabular}
\caption{(a) Surface approximation after one iteration, 
(b) element structure and (c) corresponding distance field \label{fig:sf1}}     
\end{figure}
Figure~\ref{fig:sf1} (a) shows the approximating surface after one iteration,
together with (b) the corresponding element structure and (c) the
distance field. We see that 
the domain is refined in the relevant part of the  surface.
\begin{figure}
\centering
\begin{tabular}{ccc}
(a)\includegraphics[width=4.2cm]{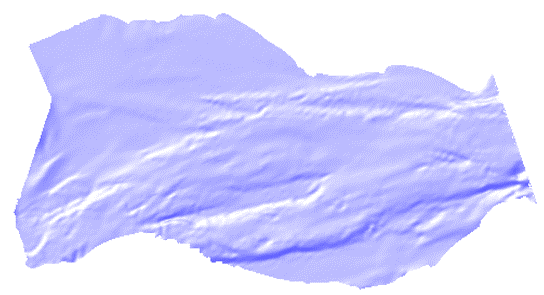}
&(b)\includegraphics[width=3cm]{images/473513_mesh4.png}
&(c)\includegraphics[width=3.4cm]{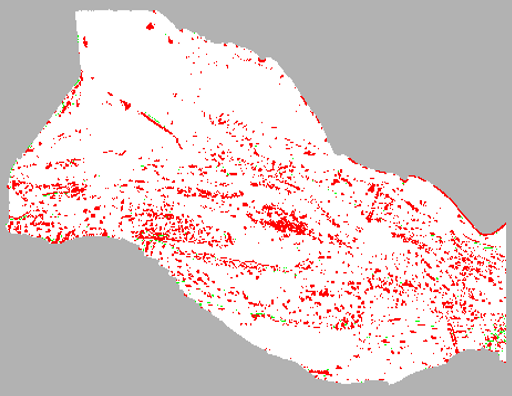}
\end{tabular}
\caption{(a) Surface approximation after four iterations, 
(b) element structure and (c) corresponding distance field \label{fig:sf4} }     
\end{figure}
After 4 iterations, it can be seen from Figure~\ref{fig:sf4} that 
the surface starts to represent
details in the sea floor. We see from the element structure that the surface has
been refined more in areas with local detail. The distance field
reveals that most of the points are within the 0.5 meter threshold.

\begin{figure}
\centering
\begin{tabular}{ccc}
(a)\includegraphics[width=4.2cm]{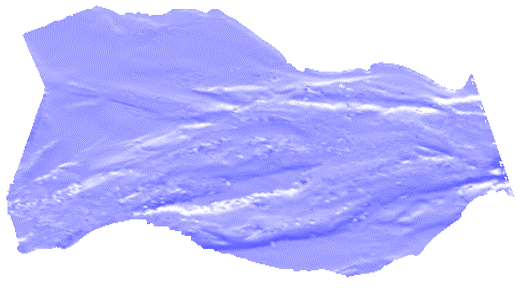}
&(b)\includegraphics[width=3cm]{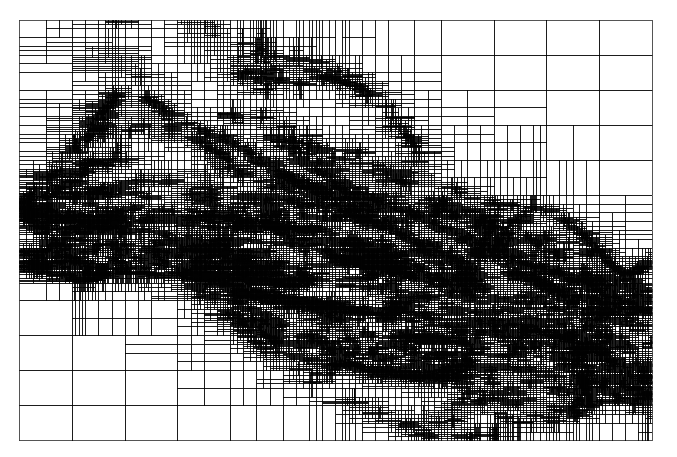}
&(c)\includegraphics[width=3.4cm]{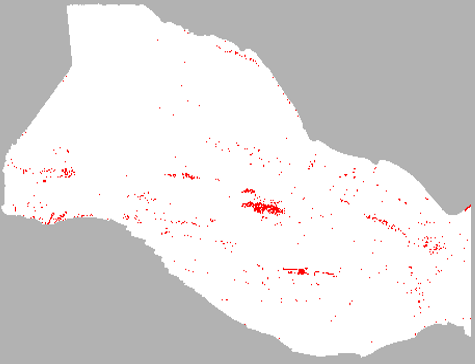}
\end{tabular}
\caption{(a) Final surface approximation after seven iterations, 
(b) element structure and (c) corresponding distance field \label{fig:sf7}}     
\end{figure}
After 7 iterations, the surface, Figure~\ref{fig:sf7} (a),
represents the shape of the sea floor very well, the corresponding
element structure (b) indicates heavy refinement in areas with local details
and only a few features in the point cloud fail to be captured by the 
surface (c). Table~\ref{fig:tab1} shows the evolution of the approximation
accuracy throughout the iterative process.

\begin{table}
  \begin{tabular}{|c|c|c|c|c|c|} \hline    
Iteration& Surface file size& No. of coefficients& Max. dist.& Average dist.& No. out points \\ \hline
0 & 26 KB & 196 & 12.8 m. & 1.42 m. & 9.9 million \\ \hline
1 & 46 KB & 507 & 10.5 m. & 0.83 m. & 7.3 million \\ \hline
2 & 99 KB & 1336 & 8.13 m. & 0.41 m. & 3.9 million \\ \hline
3 & 241 KB & 3563 & 6.1 m. & 0.22 m. & 1.4 million \\ \hline
4 & 630 KB & 9273 & 6.0 m. & 0.17 m. & 0.68 million \\ \hline
5 & 1.6 MB & 23002 & 5.3 m. & 0.12 m. & 244 850 \\ \hline
6 & 3.7 MB & 52595 & 5.4 m. & 0.09 m. & 75 832 \\ \hline
7 & 7.0 MB & 99407 & 5.3 m. & 0.08 m. & 20 148 \\ \hline
\end{tabular}
\caption{Accuracy related to approximation of a 280 MB point cloud after an
increasing number of iterations. The second and third column show the
number of coefficients in the surface and the corresponding file size.
The maximum (column 4) and average (column 5) distance between a
point and the surfaces is shown along with the number of points where the
distance is larger than 0.5 meters (column 6).\label{fig:tab1}}
\end{table}

\begin{figure}
\begin{tabular}{c c}
(a)\includegraphics[width=5cm]{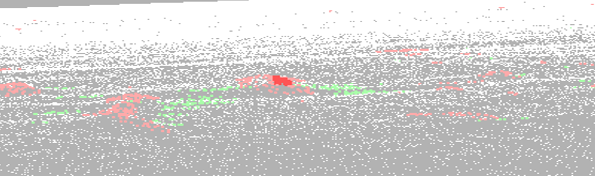} 
(b)\includegraphics[width=5cm]{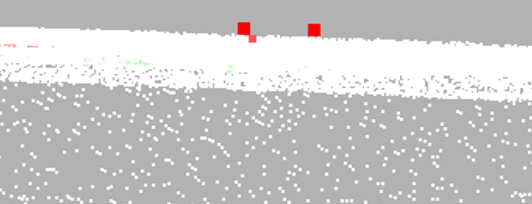}
\end{tabular}
\caption{(a) Features not entirely captured by the approximating surface, and
(b) outliers in the point set. White points lie closer to the surface than 
0.5 meters, red and green points have a larger distance. The point size and
colour strength are increased with increasing distance.  
\label{fig:distfield7}}     
\end{figure}

With every iteration, the surfaces size has increased 
while the average distance between
the points and the surface decreased, as did the number of points outside
the 0.5 meters threshold. The decrease in the maximum distance, 
however, stopped after 5 iterations. We also find that 2 points have a
distance larger than 4 meters, while 22 have a distance larger than 2 meters.
In contrast, the elevation interval is about 50 meters.
If we look into the details of the
last distance field (Figure~\ref{fig:distfield7}), we find two categories of
large distances: details that have been smoothed out (a) and outliers (b). 
If, in the first case, a very accurate surface representation is required, a 
triangulated surface should be applied in the critical areas. Outliers,
on the other hand, should be removed from the computation. Still, isolated outliers,
as in this case, do not have a large impact on the resulting surface.

{\bf Example 2} We approximate a point cloud composed from several data 
surveys taken from an area in the British channel,
and look at the result after four and seven iterations. 10 partially 
overlapping surveys contain a total of 3.2 million points. 
The accuracy threshold is again taken to be 0.5 meters.
After four iterations, the maximum distance is 27.6 meters and the average
distance is 0.2 meters. After seven iterations, the numbers are 26.9 meters and
0.08 meters, respectively. The number of points outside the threshold
are 367 593 and 38 915, respectively. Although the average approximation error
and number of points with a large distance are significantly reduced from the 
4th to the 7th iteration, the numbers are clearly poorer than for the
previous example. Table~\ref{tab:ex2} gives more detailed information.
\begin{table}
  \begin{tabular}{|c|c|c|c|c|c|c|c|c|} \hline    
Survey & No. pts & \multicolumn{3}{|c|}{4 iterations} & \multicolumn{3}{|c|}{7 iterations} & Elevation \\ \hline
&& Max bel. & Max ab. & Average & Max bel. & Max ab. & Average & \\ \hline
1 & 71 888 & -27.6 m. & 4.9 m. & 0.6 m. & -26.7 m. & 2.8 m. & 0.2 m. & 35.7 m. \\ \hline
2 & 24 225 & -8.3 m. & 6.7 m. & 0.6 m. & -5.4 m. & 4.2 m. & 0.3 m. & 27.1 m. \\ \hline
3 & 16 248 & -10.9 m. & 12.0 m. & 0.9 m. & -4.1 m. & 6.0 m. & 0.3 m. & 38.4 m. \\ \hline
4 & 483 & -1.4 m. & 6.0 m. & 0.7 m. & -1.5 m. & 4.1 m. & 0.4 m. & 11.3 m. \\ \hline
5 & 7 886 & -6.3 m. & 7.4 m. & 0.4 m. & -4.1 m. & 5.8 m. & 0.2 m. & 33.3 m. \\ \hline
6 & 4 409 & -8.3 m. & 9.2 m. & 0.5 m. & -6.1 m. & 5.6 m. & 0.2 m. & 31.6 m. \\ \hline
7 & 12 240 & -7.2 m. & 8.5 m. & 0.7 m. & -6.8 m & 9.0 m. & 0.5 m. & 30 m. \\ \hline
8 & 2 910 & -6.9 m. & 7.8 m. & 1.5 m. & -5.5 m. & 4.4 m. & 0.7 m. & 15.4 m. \\ \hline
9 & 1 049 951 & -12.7 m. & 10.5 m. & 0.4 m. & -4.2 m. & 3.1 m. & 0.1 m. & 36.1 m. \\ \hline
10 & 2 047 225 & -1.7 m. & 2.5 m. & 0.1 m. & -1.0 m. & 1.1 m & 0.06 m. & 11.9 m. \\ \hline
\end{tabular}
\caption{Approximation accuracy of the point cloud combined from 10 data surveys. 
The maximum distances below and above and the average distance
after 4 and 7 iterations are listed. The elevation range for each
data set is given for comparison. \label{tab:ex2}}
\end{table}

\begin{figure}
\begin{tabular}{cc}
\begin{minipage}{0.5\textwidth}
\centering
(a)\includegraphics[width=5.2cm]{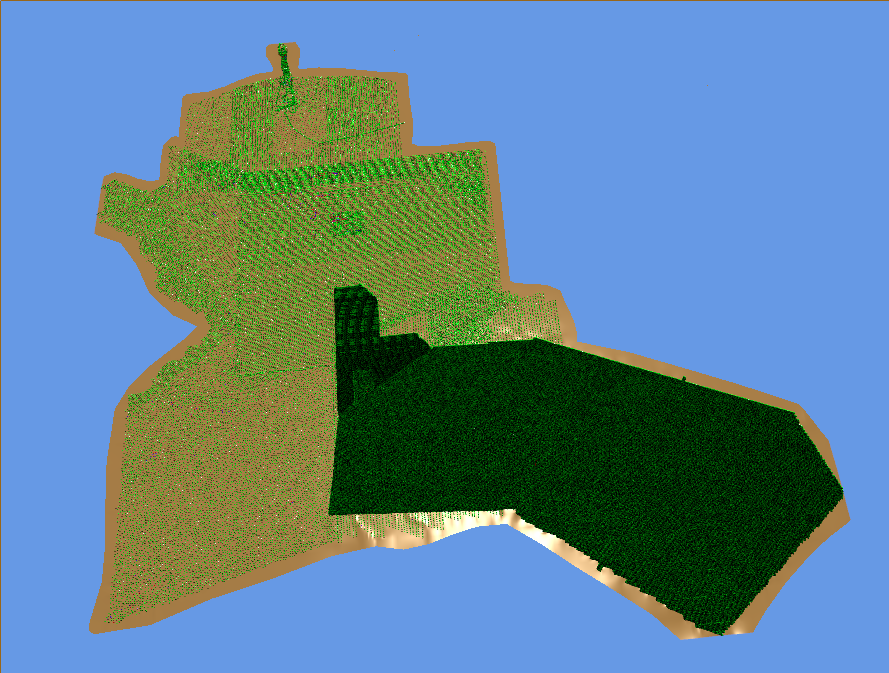}
\end{minipage}
\begin{minipage}{0.5\textwidth}
\centering
(b)\includegraphics[width=5.2cm]{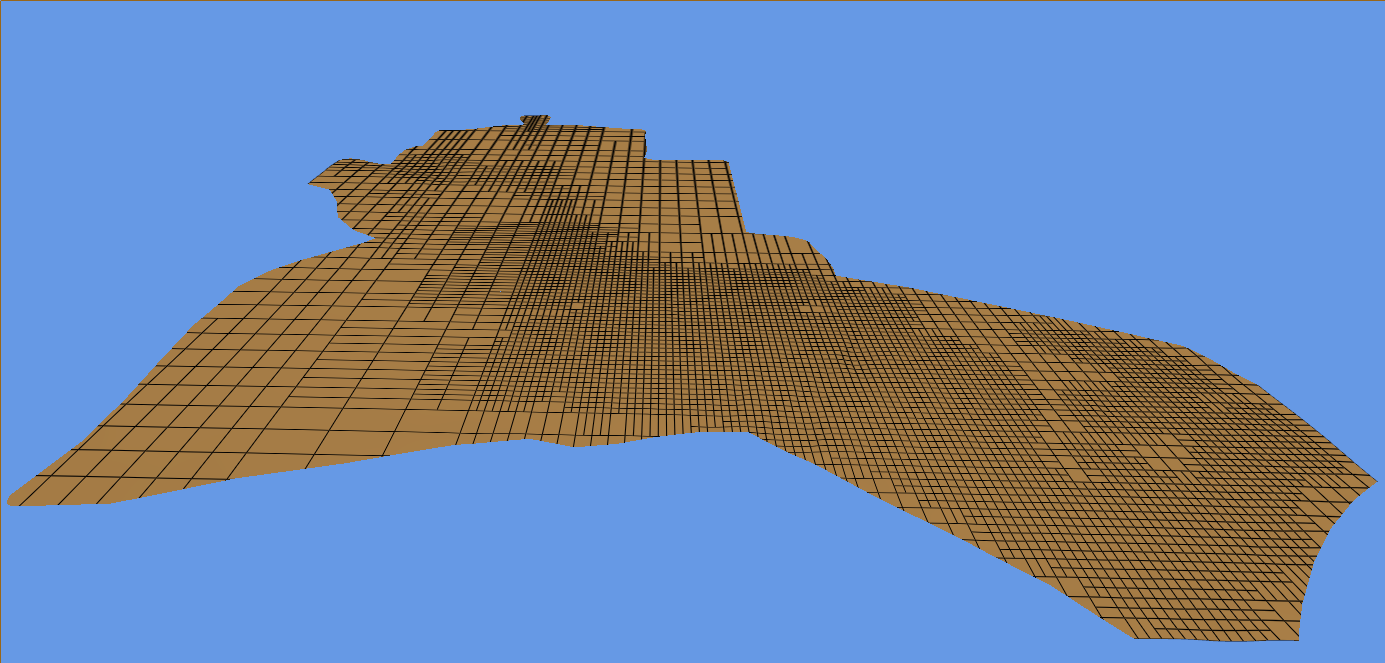} \\
(c) \includegraphics[width=5.2cm]{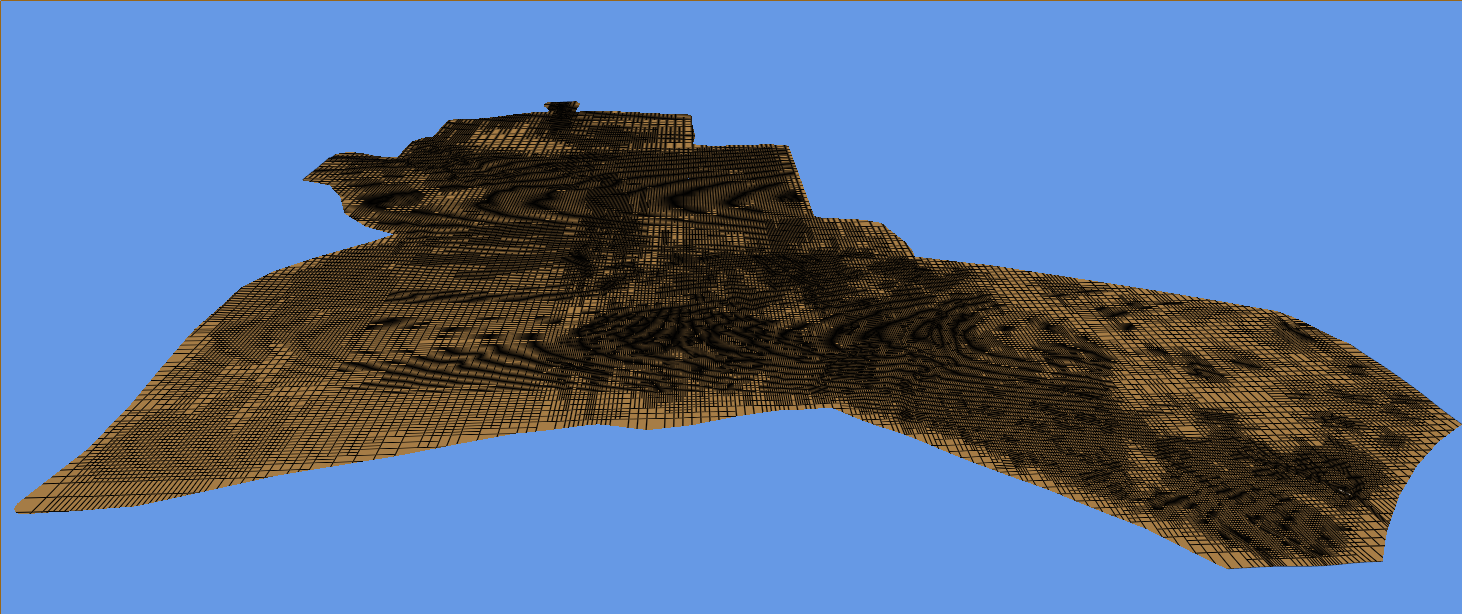} 
\end{minipage}
\end{tabular}
\caption{(a) The combined point cloud,
(b) the polynomial patches of the surface approximation after 4 iterations,
and (c) after 7 iterations. Data courtesy: SeaZone
\label{fig:example2} }
\end{figure}
Figure~\ref{fig:example2} shows the point cloud assembled from the partially
overlapping data surveys. This construction leads to a data set with a very
heterogeneous pattern, in some areas there are a lot of data points, while in others
quite few points describe the sea floor. The polynomial patches of the surface,
(b) and (c), show that the surface has been refined significantly during
the last 3 iterations.

\begin{figure}
\begin{tabular}{cc}
(a)\includegraphics[width=5.6cm]{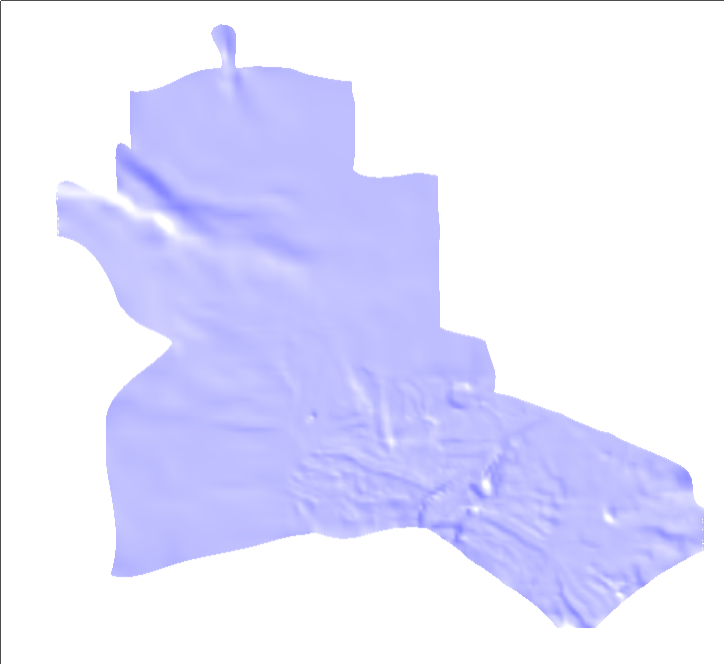}
&(b)\includegraphics[width=5.6cm]{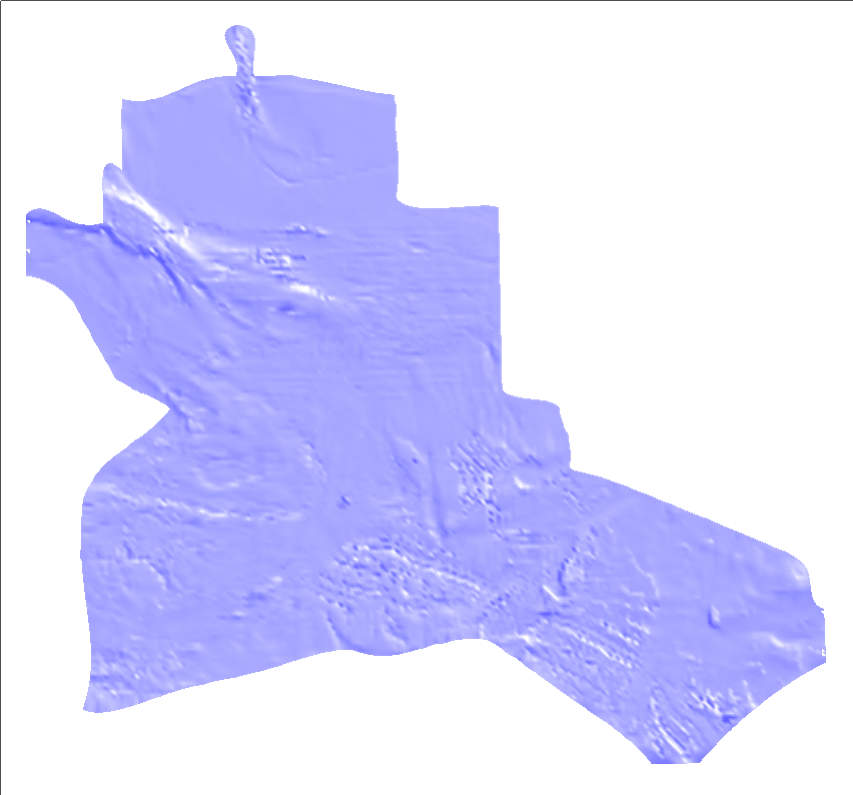}
\end{tabular}
\caption{(a) The surface after 4 iterations, and
(b) after 7 iterations
\label{fig:ex2_sf} }   
\end{figure}  
\begin{figure}
\begin{tabular}{cc}
(a)\includegraphics[width=5.8cm]{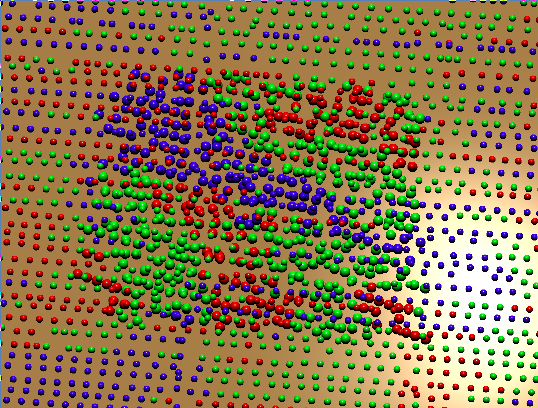}
&(b)\includegraphics[width=5cm]{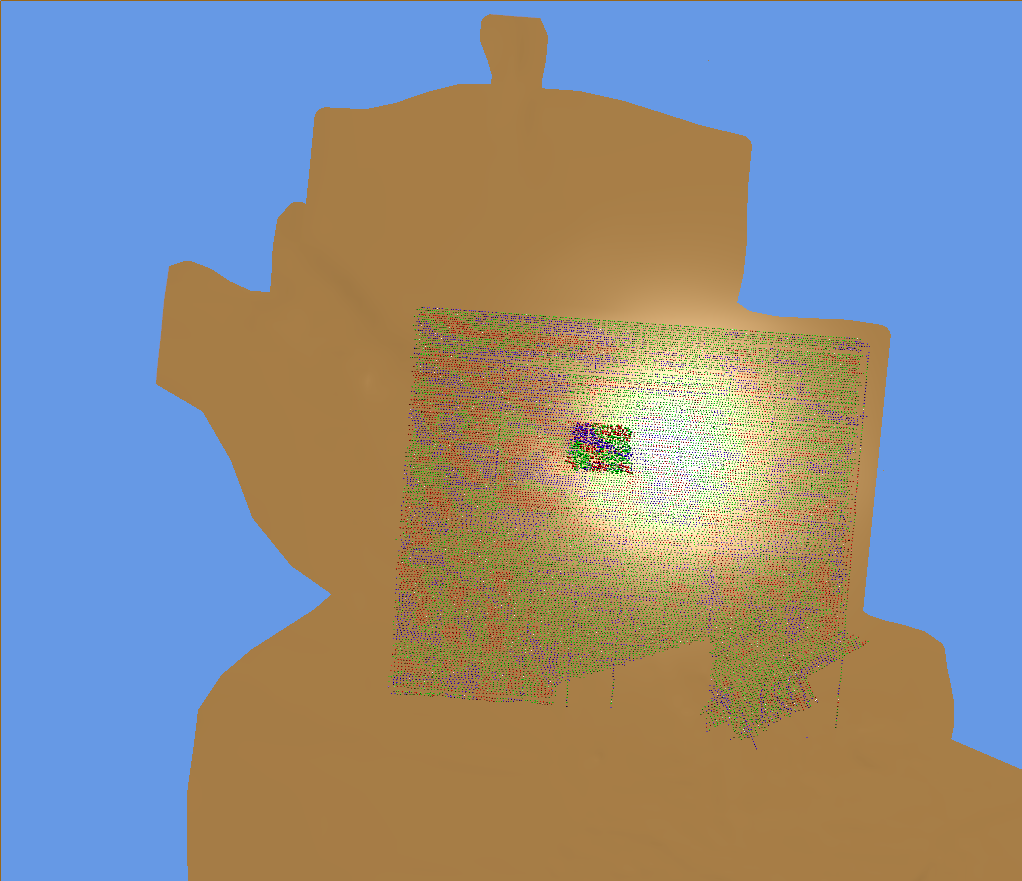}
\end{tabular}
\caption{(a) Detail of the distance field corresponding to the surface after 4
iterations for data surveys 2 and 4 in Table~\ref{tab:ex2}, distance threshold 0.5 meters, and
(b) the detail positioned in the complete surface. Green points lie closer to
the surface than 0.5 meters, while red and blue points lie outside this 
threshold on opposite sides of the surface.
\label{fig:ex2_detail1} }     
\end{figure}
Figure~\ref{fig:ex2_sf} shows the approximating
surfaces after four and seven iterations. In the first case (a), the surface
is not very accurate, as we have seen in Table~\ref{tab:ex2} and the polynomial
mesh is also quite lean, as is seen in Figure~\ref{fig:example2} (b). 
Neither, the second surface is very accurate,
but in this case some oscillations can be identified, 
Figure~\ref{fig:ex2_sf} (b), and the polynomial mesh
has become very dense; it is likely that we are attempting to model noise.

\begin{figure}
\begin{tabular}{cc}
(a)\includegraphics[width=5.4cm]{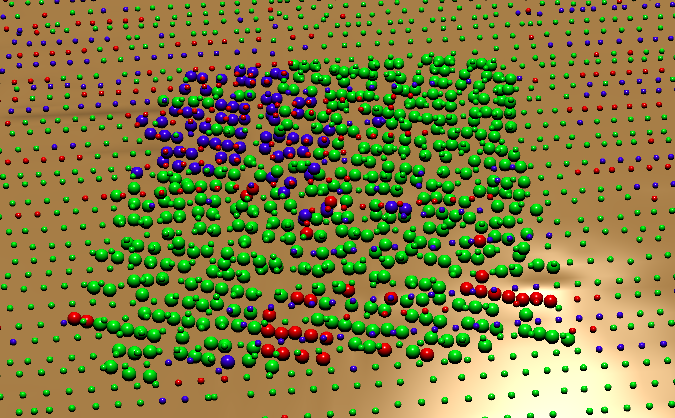}
&(b)\includegraphics[width=5.5cm]{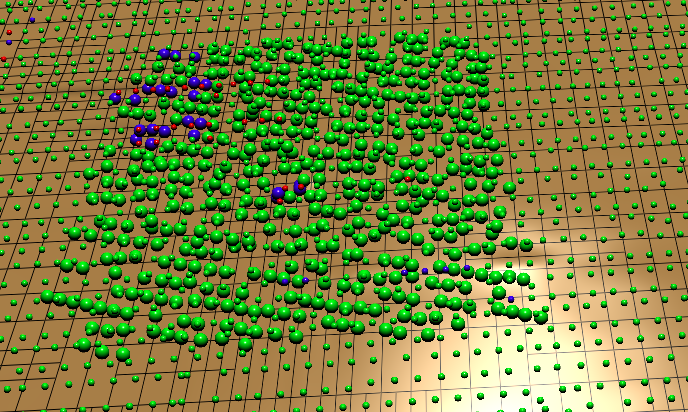}
\end{tabular}
\caption{(a) The same detail as in Figure~\ref{fig:ex2_detail1}
corresponding to the surface after 7 iterations, and 
(b) corresponding distance field with a 2 meters threshold
\label{fig:ex2_detail2} }     
\end{figure}
Figures~\ref{fig:ex2_detail1} and~\ref{fig:ex2_detail2} zoom into a 
detail on the surfaces and show the distance fields of two data surveys, 
number 2 and 4
in Table~\ref{tab:ex2}. Data set 2 is shown as small dots and
4 as large dots. In Figure~\ref{fig:ex2_detail1} (a) and 
~\ref{fig:ex2_detail2} (a), points within the 0.5
meters threshold are coloured green while red points and blue points are
outside the threshold. Red points lie below the surface and blue points
above. We see that 
points from the two data sets lie on opposite sides of the surface while
being geographically close. In
Figure~\ref{fig:ex2_detail2} (b) the distance threshold is increased to 2 meters,
and there are still occurrences where close points from the two data sets are
placed on opposite sides of the surface. Thus, the vertical distance
between these points is at least 4 meters. The polynomial elements of the
surface included in (b) indicate that a high degree of refinement has taken place
in this area. The combined data collection clearly
contains inconsistencies, and is a candidate for deconfliction

\section{Deconfliction} \label{deconfliction}
Overfitting or fitting to inappropriate data causes oscillations in the surface
and poorly reliable results. Processing the data to remove inconsistencies
and selecting the appropriate filtering criteria is a non-trivial task.
This filtering process is called deconfliction and is related to outlier detection. 

\subsection{Outlier Detection}
An outlier is an observation that is inconsistent with the remainder of the data 
set. Outlier detection is concerned with finding these observations, and as
outliers can drastically skew the conclusions drawn
from a data set, statistical methods ~\cite{statistical} for detecting 
these observations have been a topic for a long time.

Consider a data set, measurements of discrete points on the sea bottom. 
We compare the data points to a trend surface and obtain a set of residuals, 
and
want to test the hyphotesis that a given point belongs to the continuous
surface of the real sea floor. Then the corresponding residual should not be
unexpectedly large. In statistical terms, the difference surface between the
real sea bottom and our trend surface is the
population and the residual set is a sample drawn from the population. The
sample mean and standard deviation can be used to estimate the population 
mean. In order to test if a point is an outlier, i.e., not representative
of the population, we define a confidence interval. In a perfect world, 
this interval would relate to the normal distribution having zero mean and a
small standard deviation. Other distributions can, however, be more appropriate. 
For instance, the 
so called {\sl Student's t distribution} depends on the number of 
samples and is intended for small sampling sizes.

The confidence interval depends on a confidence level $\alpha$, and is given by 
$
\Big({\tilde x }-z_{\alpha /2}\frac{S}{\sqrt{n}},
{\tilde x }+z_{\alpha /2}\frac{S}{\sqrt{n}}\Big)
$.
Typically $\alpha \in [0.001, 0.2]$ and the probability that the parameter
lies in this interval is $100(1 - \alpha)$\%.
The value $z_{\alpha /2}$ denotes the parameter where
the integral of a selected distribution to the right of the
parameter is equal to $\alpha /2$. It can be
computed from the distribution, but tabulated values are also available, see
for instance~\cite{studentst} for the Student's t distribution. 
$\tilde x$ is the sample mean and
$S$ the sample standard deviation while $n$ is the number of points in the sample.

In the deconfliction setting, we want to test whether the residuals from 
different data sets can be considered to originate from the same sea floor.
I.e., we want to compare two distributions, which requires a slightly
different test.
To test for equal means of two populations, we can apply
the Two-Sample t-Test~\cite{studentst2}. To have equal means the value 
$$T=\frac{{\tilde x }_1 - {\tilde x }_2}{\sqrt(s_1^2/N_1 + s_2^2/N_2)}$$
should lie in an appropriate confidence interval. ${\tilde x }_k$ is the 
mean of sample $k, k=1,2$ and $s_k$ is the standard deviation. $N_k$ is the
number of points in the sample. If equal standard deviation is assumed
the number of degrees of freedom used to define the confidence interval 
is $N_1+N_2-1$, otherwise a more complex formula involving the
standard deviations is applied to compute the degrees of freedom. 
This test has, depending on the number
of sample points, a thicker tail than the normal distribution, but does
still assume some degree of regularity in the data. For instance, the
distribution is symmetric. Thus, we need to investigate to what extent
the test is applicable for our type of data.

Bathymetry data may contain outliers. Erroneous soundings can be caused by
several factors, including air bubbles, complexities in the
sea floor and bad weather conditions. 
These measurements need to be located and excluded from further
processing to guarantee that correct results will be generated from the cleaned data. The distinction between outliers and  data points describing real features in the sea floor is a challenge. True features should be kept and there are no firm rules saying when an outlier removal is appropriate. 

For multi beam sonars,
outlier detection is discussed in a number of papers~\cite{outliers1,
outliers2,outliers3,outliers4}. 
Traditionally outliers are detected manually by visual inspection.
However, due to the size of current bathymetry data surveys,
automatic cleaning algorithms are required.
The user can define a threshold as a multiple of the computed
standard deviation and
use statistical methods like confidence intervals or more application specific
methods developed from the generic ones to detect outliers. For instance,
Grubbs method~\cite{outliers2} is based on the Student's t distribution. 

Computations of statistics for outlier removals may be based on the depth 
values themselves, but often residuals with respect to a trend surface are
preferred. In the latter case, the trend surface is typically computed for
subsets of the data survey. Selecting the cell size for such subsets
is non-trivial. Large cells give larger samples for the computation of
statistical criteria, but on the other hand, the cells size must be limited
for the trend surface to give a sufficiently adequate representation of
the sea floor.
In~\cite{outliers4} a multi-resolution strategy is applied to get a reasonable
level of detail in the model used for outlier detection.
The selection of a suitable neighbourhood of interest for an outlier is
relevant also for other types of outlier detection algorithms, for instance 
proximity based techniques as in k-Nearest Neighbour 
methods~\cite{outliers2}.
A problem in trend surface analysis is that the surface
tends to be influenced by the outliers. It has been proposed~\cite{outliers5} 
to minimize this
influence by using a minimum maximum exchange
algorithm (MMEA) to select the data points for creating the trend surface.
In~\cite{outliers3}, the so called M-estimator is utilized for the
surface generation.

\subsection{Preparing for Deconfliction}
Deconfliction becomes relevant when we have more than one data survey 
overlapping in a given area. Two questions arise:
are the data surveys consistent, and if not, which survey to choose? The first
question is answered by comparing statistical properties of the data
surveys. The answer to the second is based on properties of each data
survey. The data surveys are equipped with metadata information. This
includes the acquisition method, date of acquisition, number of points and point
density. Usually, the most recent survey will be seen as the most reliable,
but this can differ depending on the needs of the application, for instance when
historical data is requested. In any case, an automated procedure is applied
for prioritizing the data surveys resulting in scores that allow, at
any sub-area in the region of interest, a sorting of overlapping surveys. 
We will not go into details about the prioritization algorithm.

In the first surface generation example, we observed a couple of outliers that
could be easily identified by their distance to the surface.
Considering outlier data sets, we want to base the identification on
residuals to a trend surface, also called reference surface.
In~\cite{outliers3} low order polynomials approximating hierarchical data
partitions defined through an adaptive procedure were used as trend surfaces.
We follow a similar approach by choosing an LR B-spline surface as the
trend surface and use the framework described in Section~\ref{sec:adaptive}
to define a surface roughly approximating the point cloud generated by
assembling all data surveys. 

The deconfliction algorithm is applied for each polygonal patch in the surface.
This patch will, in the following be called element, and the
element size has a significant impact on the result.
Too many degrees of freedom compared to the number of data points results in
the reference
surface modeling the anticipated noise in the data, while too few will lead to
a situation where the statistical properties derived from the residuals
become less trustworthy.
The strategy for adaptive refinement of an LR B-spline
surface implies that the surface will be refined in areas where the 
accuracy is low. Thus,
the size of the polynomial elements will vary: in regions where there is a lot
of local detail, the element size will be small, while in smooth regions or
regions where the point density is too low to represent any detail, the element
size is large. Example 1 in Section~\ref{surfgenexample} shows the element
mesh for an LR B-spline surface at different iteration levels. 
Adaptive refinement automatically implies an adaptive size of the
surface elements. However, the number of iterations performed
in the algorithm must be selected to get a good basis for the decisions, see
Section~\ref{sec::deconfexamples}, Figure~\ref{fig:sf0} to~\ref{fig:sf7} for 
an example of the effect of the refinement level of the reference surface.

\subsection{The Deconfliction Algorithm}
Outliers appear to be inconsistent with the general trend of the data.
It is in the nature of outlier detection that there is a subjective
judgement involved.
Our aim is to develop an automatic outlier detection algorithm where
the outliers are subsets of data surveys and where the sample pattern is
extremely non-uniform. 

If more than two surveys overlap in a domain they are tested pairwise with
respect to score. The second highest scored survey is first compared
to the one with highest score. Every new survey is tested against
all previously accepted surveys and needs to be found consistent with all to be accepted.

After applying the deconfliction, the cleaned data surveys are used to update
the reference surface to obtain a final surface with better accuracy. This
is done by the surface generation algorithm described in Section~\ref{surfgen},
but the process is started from the reference surface and not from a lean
initial tensor-product spline surface. Thus, fewer iterations are required to
obtain a sufficient accuracy.

Suppose two or more data surveys overlap in an identified area. 
The point cloud assembled
from all the surveys is approximated by an LR B-spline surface of
low accuracy. The consistency check is performed elementwise and pairwise.
For one comparison
of two surveys, several aspects must be taken into consideration:
\begin{itemize}
\item The pattern described by the combined data surveys may be very 
non-uniform.
\item The number of points within an element may differ greatly from
element to element and from survey pair to survey pair.
\item The data surveys may cover roughly the same area within an element, they
may be completely disjoint or overlap in a tiny area.
\item The number of points in each survey may differ by an order of magnitude. 
The data size of a survey is independent of its priority score. 
The reference surface will favour the survey with many points.
\item One or both data surveys may contain outliers.
\item If the two surveys have  the same score and overlap barely or not at 
all, this probably implies that the surveys originate from the same
acquisition, but the point set is split at some stage. This is treated as a 
special case.
\end{itemize} 

Given this survey configuration, can methodologies from statistics or
multi beam outlier detection apply? Given subsets from two
data surveys, we want to determine if they belong to the same underlying
surface.
The following properties are taken into account in the algorithm:
\begin{itemize}
\item The mean of the two samples and the difference between these means.
\item The range of distances to the reference surface for each sample.
\item The standard deviation of the signed distances between the
sample points and the reference
surface for each sample and for the data set obtained by combining 
the two samples.
\item Size of overlap between the sample domains relative to the maximum
sample domain size.
\item The Two sample t-Test value and the associated confidence interval.
\end{itemize}
A sample in this context is a data survey restricted to one surface element.
An immediate observation is that the Two sample t-Test is very strict 
for this kind of data and that the value becomes very large when the
standard deviations of the two samples are small. Thus, applying this test
directly would be too strict. However, the t-Test value tends to vary
consistently with the other properties. When this tendency is contradicted
a closer investigation should be initiated. Similarly, if the standard
deviation of one or both data surveys is large, this indicates outliers
within the data sets or a high degree of detail in the sea bottom. Also 
in this case more testing can be beneficial and the deconfliction test
is applied to sub-domains within the element.

The surveys are considered consistent if the following criteria hold:
\begin{itemize}
\item The sample means are close relative to the surface generation 
threshold.
\item The range of the distances field of the candidate sample does not exceed the
range of the high priority sample with more than an amount 
deduced from this threshold.
\item Most of the distances computed from points from the candidate sample
lies within the range of the prioritized sample.
\item The standard deviation computed from the combined data set does not exceed the
individual standard deviations with more than a small fraction.
\end{itemize}
If some of the conditions above do not apply, but the overlap between 
the samples
is small, the test is repeated on a sub-domain where there is a significant
overlap between the samples.
The possibility of consistency
checking on reduced domains implies a second level of adaptivity in
addition to the adaptivity in creating the reference surface, even though
the reference surface is not updated. 

Surface generation, even with a careful selection of approximation method,
is sensitive to patterns in the data points. 
Empty regions with significant variation in the height values may lead to
unwanted surface artifacts. However, even if one data survey lacks points in 
an area, another survey may contain information about the area. Thus, the
combination of several surveys can give more complete information than one
survey alone, as long as the information from the different surveys is
consistent. The question is: What should be done with a group of points that
are found to be inconsistent or possibly inconsistent with the remainder of
the points in the area? Is it more damaging for the final surface to keep them
or remove them? The answer depends on the configuration of points. If the 
candidate outliers are disjoint from the higher prioritized point clouds,
and the distance between the point clusters large enough to fit a reasonable
surface, the group of candidate outliers should be kept. Otherwise, the
points should be removed.

In the following, we will look into a couple of different classes of
configurations and discuss them in some detail. The algorithm classifies
sub point clouds into consistent, not consistent and indeterminate, based on
statistics on the distance field.
The indeterminate cases are first investigated in more detail using the
sub-element approach mentioned above and if the case is still classified
as indeterminate, treated again using knowledge on 
other elements covered by the same data surveys, to tune
the algorithm.

\begin{figure}
\begin{tabular}{c c c} 
a) \includegraphics[width=3cm]{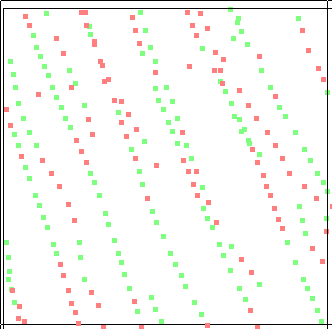}
&b) \includegraphics[width=3cm]{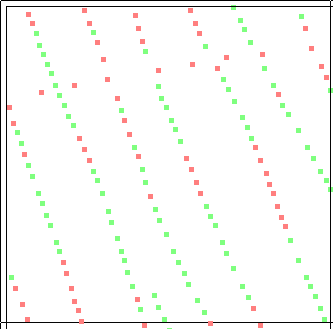}
&c) \includegraphics[width=3cm]{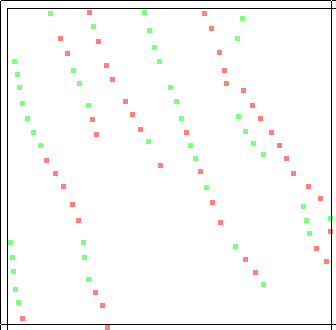}
\end{tabular}
\caption{(a) Pattern of residuals for both surveys, (b) high prioritized survey and
(c) survey of lower priority, Element Example 1. Red points lie above the 
reference surface and green points below.
\label{fig:el1}}       
\end{figure}
{\bf Element Example 1} We look at is a detail in the test case covered in the
first example of Section~\ref{sec::deconfexamples}. The element is
overlapped by two of the data surveys, and the patterns of the two data 
surveys are relatively similar as seen in Figure~\ref{fig:el1}. 

\begin{table}
\begin{center}
\begin{tabular}{|c|c|c|c|c|c|c|} \hline 
Survey & Score & No pts & Range & Mean & Std dev & Size  \\ \hline
1 & 0.657 & 152 & -0.232, 0.250 & -0.021 & 0.0088 & 1863.9 \\ \hline
2 & 0.650 & 86 & -0.155, 0.172 & -0.003 & 0.0046 & 1823.0 \\ \hline
\end{tabular}
\caption{Characteristic numbers for residuals, the reference surface
is created with 3 iteration levels. Element Example 1.
\label{tab:el1_1}}       
\end{center}
\end{table}
The range of the distance field at 3
iterations, the mean distance standard deviation and domain size for the two surveys are
given in Table~\ref{tab:el1_1}. The domain sizes are given as the bounding
box of the $x-$ and $y-$ coordinates of the points.
The overlap between the surveys has size 1802.3, 
which imply almost full overlap. The standard deviation computed from the 
combined point clouds is 0.007.
The Two sample t-Test value is 20.5 while the limit with 
$\alpha = 0.025$ is 1.96. 
The range and standard deviation for the low priority data surveys is lower
than for the prioritized one. The differences between range extent and mean value for the
two surveys are small compared to the threshold of 0.5 and the standard 
deviation doesn't increase when the two surveys are seen as one unity. Thus,
the surveys look quite consistent even if the T-test value is high
compared to the confidence interval, and
this is indeed the conclusion of the test.

\begin{figure}
\begin{tabular}{c c c c} 
a) \includegraphics[width=3cm]{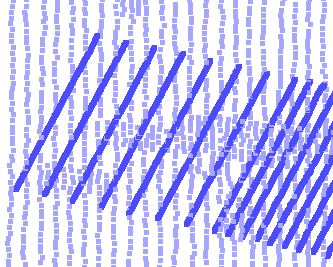}
&b) \includegraphics[width=2.45cm]{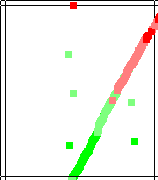}
&c) \includegraphics[width=2.45cm]{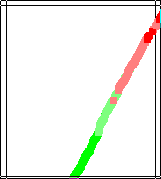}
&d) \includegraphics[width=2.45cm]{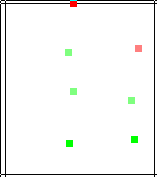}
\end{tabular}
\caption{(a) Overlapping data surveys, (b) residuals pattern for both surveys
restricted to one element, (c) prioritized survey and
(d) survey to be tested, Element Example 2. Red points lie above the 
reference surface and green points below.
\label{fig:el2}}       
\end{figure}

{\bf Element Example 2} The next example, see Figure~\ref{fig:el2}, is taken from an area with two
overlapping surveys of different patterns. The one with highest score
consists of scan lines where the points are close within one scan line, but the
distances between the scan lines are large. For the other survey, the 
points are more sparse, but also more regular. In this configuration, we 
would prefer to keep most of the points between the scan lines, but only as long as they
are consistent with the scan line points.

\begin{table}
\begin{center}
\begin{tabular}{|c|c|c|c|c|c|c|} \hline 
Survey & Score & No pts & Range & Mean & Std dev & Size  \\ \hline
1 & 0.640 & 172 & -1.05, 0.625 & -0.191 & 0.177 & 3045.3 \\ \hline
2 & 0.576 & 7 & -0.64, 1.19 & -0.028 & 0.326 & 2435.9 \\ \hline
\end{tabular}
\caption{Characteristic numbers for residuals, deconfliction level 3. 
Element Example 2.
\label{tab:el2}}       
\end{center}
\end{table}
The mean values of the residuals quite similar, see Table~\ref{tab:el2}, 
but the ranges don't
overlap well, which indicates a rejection of the survey with the lower
score. However, the individual standard deviations are relatively high, in particular
for the second survey. Thus, a more detailed investigation is initiated.
\begin{table}
\begin{center}
\begin{tabular}{|c|c|c|c|c|c|c|c|} \hline 
Sub-domain & Survey & Score & No pts & Range & Mean & Std dev & Size  \\ \hline
1 & 1 & 0.640 & 12 & -0.96, -0.56 & -0.65 & 0.015 & 13.1 \\ \hline
1 & 2 & 0.576 & 2 & -0.64, -0.24 & -0.44 & 0.040 & 44.8 \\ \hline
2 & 1 & 0.640 & 87 & -1.05, 0.10 & -0.48 & 0.062 & 698.4 \\ \hline
2 & 2 & 0.576 & 2 & -0.54, -0.15 & -0.35 & 0.039 & 35.9 \\ \hline
3 & 1 & 0.640 & 73 & -0.26, 0.62 & 0.22 & 0.035 & 597.1 \\ \hline
3 & 2 & 0.576 & 1 & 0.27, 0.27 & 0.27 &  &  \\ \hline
3b & 1 & 0.640 & 22 & 0.18, 0.37 & 0.30 & 0.004 & 35.8 \\ \hline
3b & 2 & 0.576 & 1 & 0.27, 0.27 & 0.27 &  &  \\ \hline
\end{tabular}
\caption{Characteristic numbers for residuals, sub-domains of Element Example 2
\label{tab:el2_2}}       
\end{center}
\end{table}
In sub-domain 1, the combined standard deviation is 4.75, which is way 
above the standard deviations for the individual sub surveys. However, the 
sub surveys don't overlap and after looking into the closest situated points
in the two surveys, the conclusion is that the surveys are consistent. In sub-domain 2, the
combined standard deviation is 0.537 and there is no overlap between the
two sub surveys. The conclusion is consistence for the same reason as for the 
previous sub-domain. In sub-domain 3, the combined standard deviation is 0.85.
The single point from Survey 2 is well within the range of Survey 1, but
the standard deviation tells a different story. However, after limiting the
domain even more to cover just the neighbourhood of the survey 2 point, the
characteristic residual numbers can be seen in Table~\ref{tab:el2_2} as 
sub-domain 3b and the combined standard deviation is 0.003. The survey is
accepted also in this domain. In the last sub-domain, Survey 1 has no points
and the final conclusion is acceptance.

\subsection{Deconfliction Examples} \label{sec::deconfexamples}
\begin{figure}
\begin{tabular}{c c} 
a) \includegraphics[width=5cm]{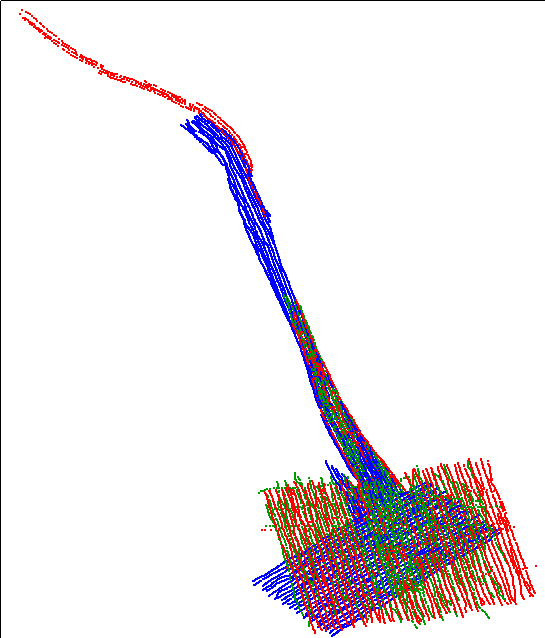}
&b) \includegraphics[width=6.5cm]{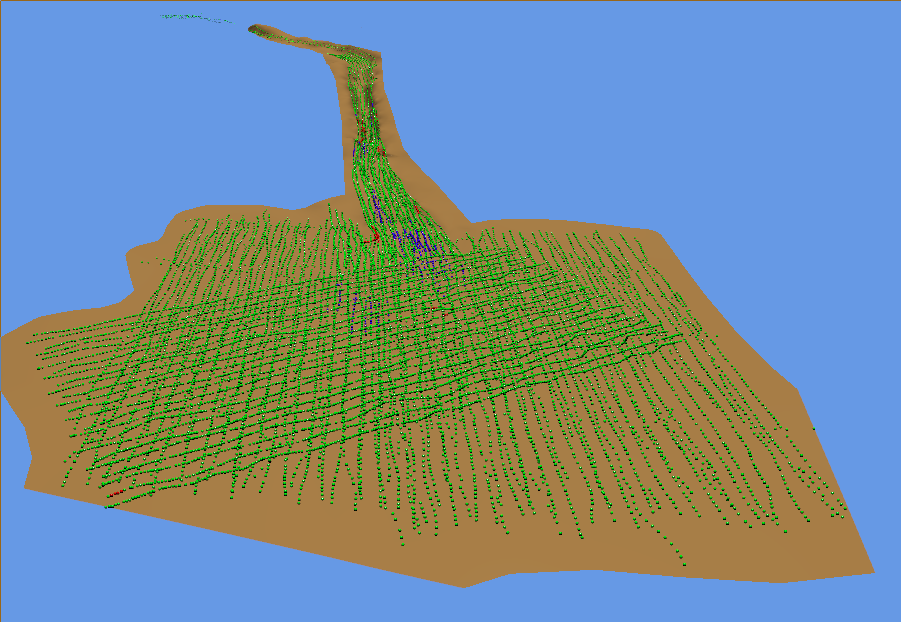}
\end{tabular}
\caption{(a)Three overlapping data surveys and (b) the combined point cloud 
with the final approximating surface. Data courtesy: SeaZone \label{fig:deconf4pts}}
\end{figure}

{\bf Example 1} Our first example is a small region with three overlapping data surveys,
Figure~\ref{fig:deconf4pts} a. The red one (survey 1 in Table 
~~\ref{tab:reflevel}) has priority score 0.675, the green (survey 2)
has score 0.65 and the blue (survey 3) 0.097. 

\begin{figure}
\begin{tabular}{c c} 
a) \includegraphics[width=5.7cm]{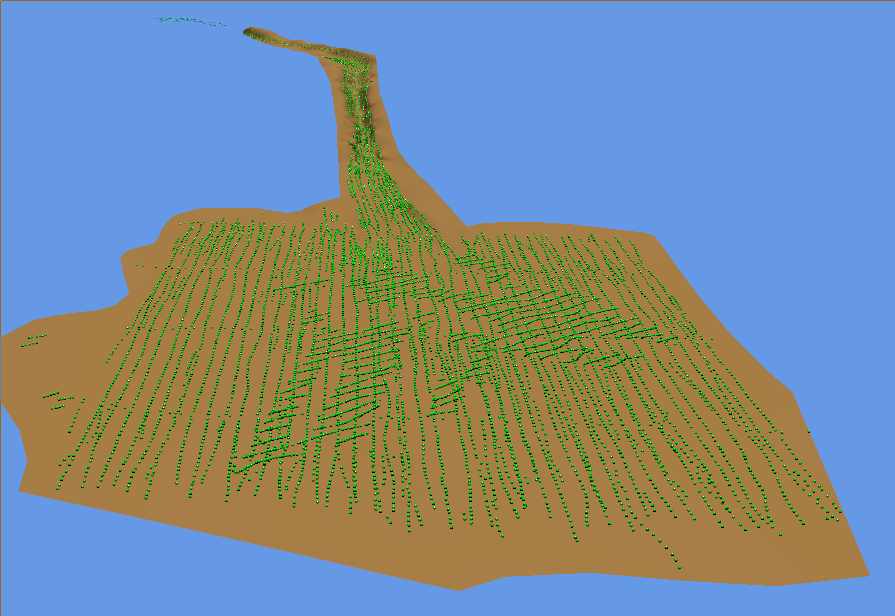}
&b) \includegraphics[width=5.7cm]{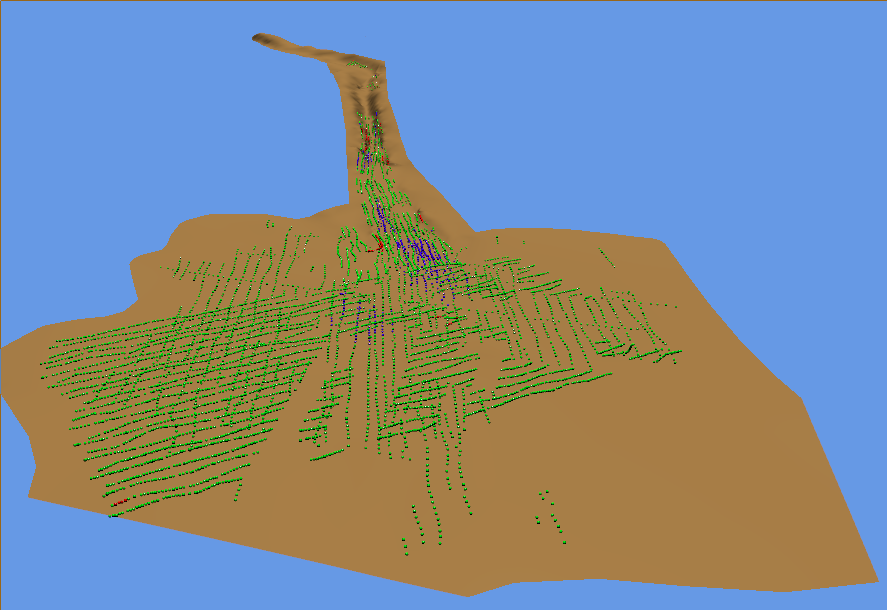}
\end{tabular}
\caption{Surface approximation and  (a) the cleaned
point set and (b) the points removed by the deconfliction. Green points lie
closer to the reference surface than the 0.5 meter threshold, red points lie
below the surface and blue points lie above, both groups lie outside the threshold.
\label{fig:deconf4res}}       
\end{figure}
The combined data set is approximated by a reference surface using 4
iterations of the adaptive surface generation algorithm. Deconfliction is
applied and the surface generation is continued, approximating only the
cleaned point set for 3 more iterations. The result can be seen in 
Figure~\ref{fig:deconf4res}. About half the points are removed by the
deconfliction algorithm and almost all the cleaned points are within the
prescribed threshold of 0.5 meters of the final surface. The points that have 
been removed from the computations, are more distant. However, most of them
are also close to the surface. In most of the area, the sea floor is quite 
flat and even if the data surveys are not completely consistent,
the threshold is quite large. In the narrow channel at the top of the data
set, the shape becomes more steep and the difference between the cleaned
and the remaining points becomes larger.
\begin{figure}
\begin{tabular}{c c} 
a) \includegraphics[width=5.7cm]{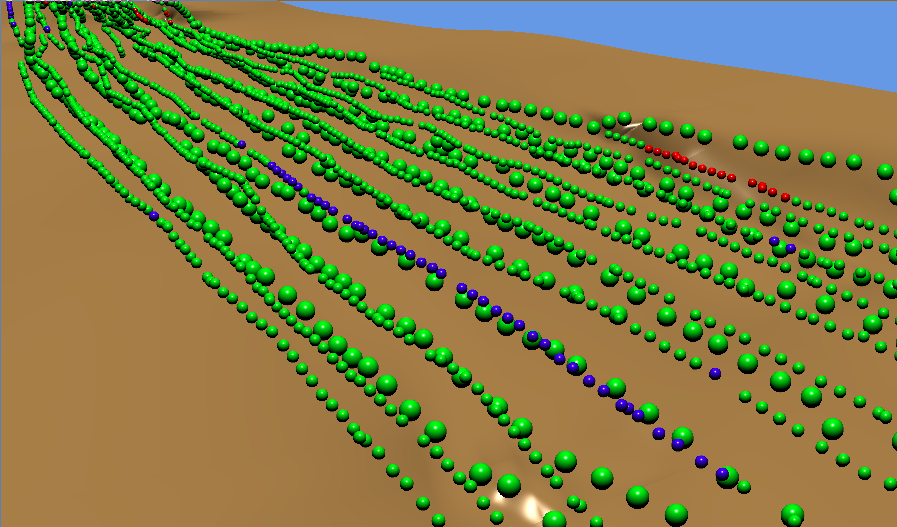}
&b) \includegraphics[width=5.7cm]{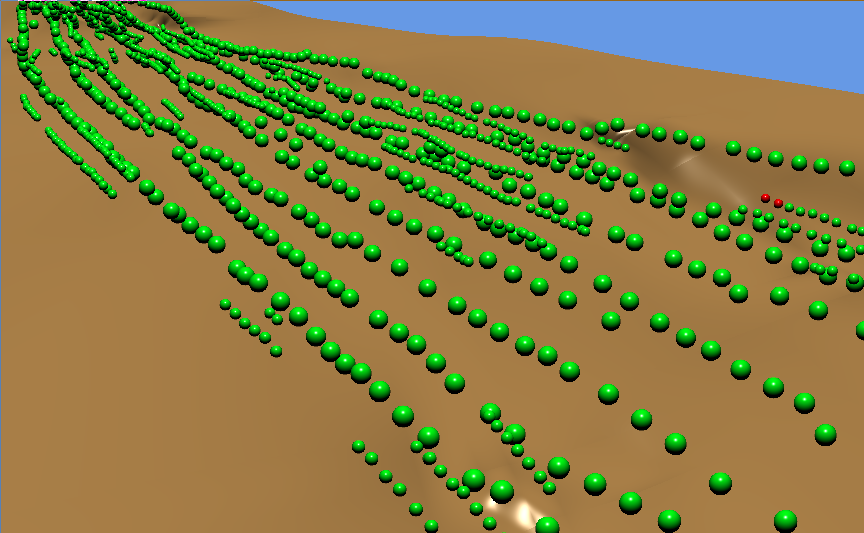}
\end{tabular}
\caption{A detail with data survey nr 2 and 3, (a) both surveys and
(b) only the highest prioritized one
\label{fig:deconf4detail}}       
\end{figure}
Figure~\ref{fig:deconf4detail} shows a detail close to the channel. In 
Figure~\ref{fig:deconf4detail} (a) two surveys are shown, 
and the one with large points has highest 
priority score. For the other one, some points lie outside the 0.5 meters
threshold (blue points), and we can see that the corresponding
scan line has different behaviour vertically than the nearby completely 
green scan line of the high priority survey.

\begin{table}
  \begin{tabular}{|c|c|c|c|c|c|c|c|c|c|c|} \hline    
\multicolumn{2}{|c|}{Survey} & No. pts & \multicolumn{2}{|c|}{No deconfliction} & \multicolumn{3}{|c|}{Deconfliction at level 3} & \multicolumn{3}{|c|}{Deconfliction at level 4} \\ \hline
\multicolumn{2}{|c|}{}&& Range & mean  & Range & mean & no. pts & Range & mean  & no. pts \\ \hline
1 & all & 6333 & -0.83, 0.70 & 0.12 & -0.49, 0.52 & 0.10 && -0.48, 0.56 & 0.09 & \\ \hline
1 & clean & && & -0.48, 0.52 & 0.10 & 6333 & -0.48, 0.56 & 0.09 & 6333 \\ \hline
2 & all & 3811 & -0.64,0.70 & 0.15 & -1.03, 1.75 & 0.21 && -0.89,1.8 & 0.20 & \\ \hline
2 & clean & && & -0.39, 0.46 & 0.10 & 1478 & -0.42,0.50 & 0.10 & 1546 \\ \hline
3 & all & 11364 & -0.55, 0.56 & 0.10 & -1.43, 1.50 & 0.18 && -1.38,1.66 & 0.18 & \\ \hline
3 & clean & && & -0.6, 0.5& 0.10 & 5209 & -0.49, 0.48& 0.10 &5430 \\ \hline
\end{tabular}
\caption{Comparison with different levels of approximation for the reference
surface
\label{tab:reflevel}}
\end{table}
Table~\ref{tab:reflevel} shows how the choice of refinement levels for the
reference surface influences the accuracy of the final surface, when 3 and
4 iterations for the reference surface is applied. For 
comparison, the surface approximation is performed also on the combined
points set without any deconfliction. The surveys are prioritized according to 
their number, and the distance range and mean distance to the reference
surface is recorded for all computations in addition to the total number
of points for each data survey and the number of points in the cleaned
survey after deconfliction. All distances are given in meters. 
In total, for the final surface, the number of iterations is
7 in all cases, but the data size of the final surfaces differ: The
surface generated without any deconfliction is of size 329 KB, the surface
with deconfliction level 3 is 131 KB large while the deconfliction level 4
surface is of size 147 KB. The distances between the final surface and
the cleaned point clouds are slightly larger, and some more points are
removed when deconfliction is performed at iteration
level 3, but the accuracy weighed against surface size is more
in favour of this choice of deconfliction level. The distances when no
deconfliction is applied are larger when compared to the numbers for the
cleaned point clouds, but smaller when all points are taken into account. 
This is no surprise as in the other case, only the cleaned point sets were
used for the last iterations of the surface generation. The numbers don't
clearly favour either deconfliction level 3 or 4. They are roughly comparable,
but the reduced surface size for level 3 is preferable.

\begin{figure}
\begin{tabular}{c c} 
a) \includegraphics[width=5.7cm]{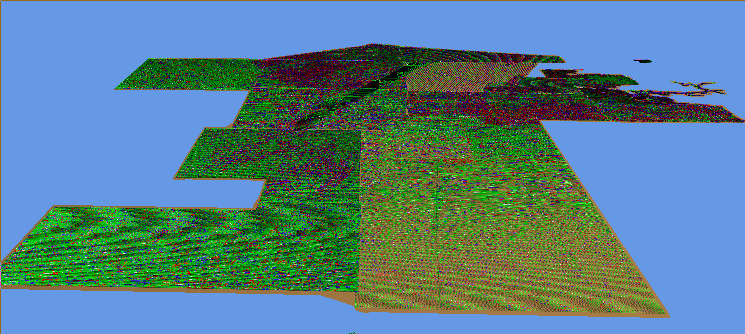}
&b) \includegraphics[width=5.7cm]{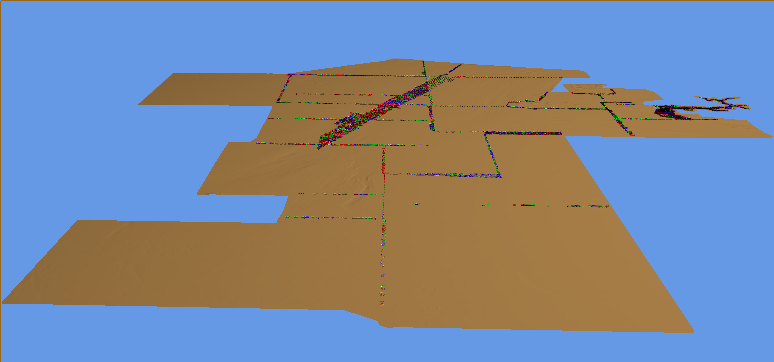}
\end{tabular}
\caption{The reference surface with (a) the points kept by the
deconfliction and (b) the points removed. Distances are computed
with respect to the reference surface, green points lie closer than 0.5
meters, red points lie below and blue points above. Data courtesy: SeaZone
\label{fig:deconex2-1}}       
\end{figure}

{\bf Example 2} This example is of a different magnitude. 255 data surveys sum up
to 1.5 GB. The data set is split
into $5 \times 3$ tiles and are approximated by surfaces.
As we can see in Figure~\ref{fig:deconex2-1}, there is limited overlap between
the data surveys. 

\begin{figure}
\begin{tabular}{c c} 
a) \includegraphics[width=5.7cm]{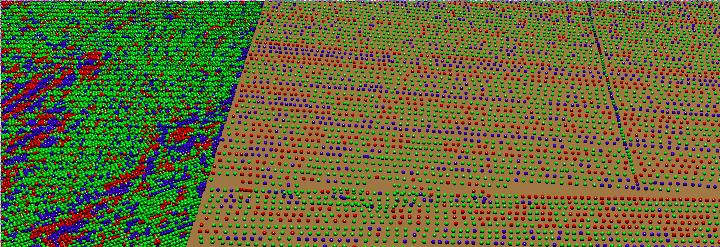}
&b) \includegraphics[width=5.7cm]{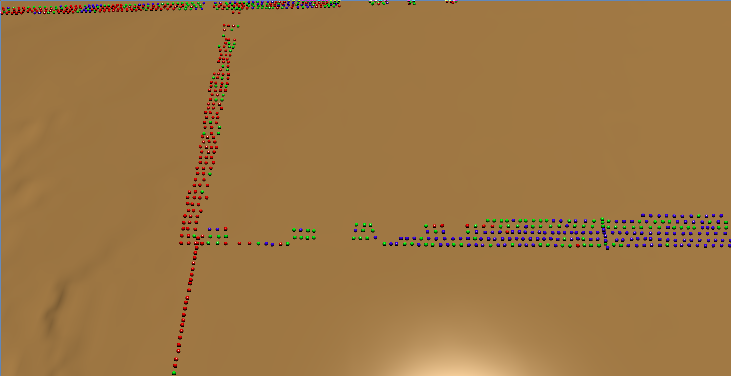}
\end{tabular}
\caption{A detail of the reference surface with (a) the points kept by the
deconfliction and (b) the removed points. 
\label{fig:deconex2-2}}       
\end{figure}
Figure~\ref{fig:deconex2-2} shows overlap zones between three data surveys
together with the kept points (a) and the removed points (b). The distances
are computed with respect to the reference surface, which is made
with deconfliction level 4. The point
colours in these zones indicate that the points from different surveys
are more than twice the tolerance apart, and consequently the overlap
points from the lowest prioritized survey are removed. %

\section{Conclusion and Further Work} \label{conclusion}
A good data reduction effect has been obtained by approximating bathymetry
point clouds with LR B-spline surfaces. The approach handles
inhomogeneous point clouds and can be used also
for topography data, but is mostly suitable if the data set is to some extent
smooth or if we want to extract the trend of the data. Data sets that mainly
represent vegetation are less suitable.

We have developed an algorithm for automated deconfliction given a set of
overlapping and possibly inconsistent data surveys. The cleaned point sets
lead to surfaces with a much smaller risk of oscillations due to noise in
the input data. The results so far are promising, but there is still 
potential for further improvements. Interesting aspects to investigate include:
\begin{itemize}
\item Outlier removal in individual data surveys prior to deconfliction.
\item Investigation of secondary trend surface approximations based on residuals
in situations with many points in an element and small overlaps between the data sets,
to detect if there is a systematic behaviour in the approximation errors
with respect to the current reference surface.
\item Continued investigation of the effect of refinement of the
LR B-spline surface to create a suitable reference surface. Aspects to study
are number of iterations and a possibility for downwards limitations 
regarding element size and number of points in an element.
\item There is no principal difference between surface modelling and
deconfliction in 2.5D and 3D. Still, an investigation regarding which
dimensionality to choose in different configurations could be useful.
\item A data survey can be subject to a systematic difference with respect
to another survey due to differences in registration, for instance the
vertical datum can differ. Identification and correction of such occurrences
are not covered by the current work. Differences in registration is a global 
feature of the data set. Indications of it can be detected locally for the
reference surface elements, but the determination of an occurrence must be made
globally.
\end{itemize}

\end{document}